\documentclass[aps,prx,superscriptaddress,twocolumn,10pt]{revtex4-2}
\usepackage[utf8]{inputenc}
\usepackage[english]{babel}

\usepackage[titletoc,toc,title,page]{appendix}
\usepackage{csquotes}

\usepackage{microtype} 

\usepackage{bm} 

\usepackage{dsfont} 
\usepackage{amsmath,amssymb,amsthm,thmtools}
\usepackage{mathtools}
\usepackage{cases}
\usepackage{calc}
\usepackage{mathrsfs} 
\usepackage[normalem]{ulem} 
\newcommand\redsout{\bgroup\markoverwith{\textcolor{red}{\rule[0.5ex]{2pt}{1.pt}}}\ULon}
\usepackage{parskip}
\usepackage[colorlinks=true]{hyperref}
\usepackage{nameref}
\usepackage[nameinlink]{cleveref}
\crefname{appsec}{Appendix}{Appendices}
\crefname{box}{Box}{Box}
\hypersetup{
  colorlinks   = true, 
  urlcolor     = green!80!black, 
  linkcolor    = blue, 
  citecolor    = red!80!black 
}
\usepackage{amsmath}
\usepackage{physics} 
\usepackage{float} 
\usepackage{graphicx}
\usepackage[usenames,dvipsnames,table]{xcolor}
\usepackage{easyReview}
\usepackage{tikz}

\usetikzlibrary{calc,shapes.geometric}

\usepackage{placeins}
\usepackage{float}
\usepackage{multirow,tabularx,booktabs}
\usepackage[most]{tcolorbox}
\newtcbtheorem{tbox}{Box}{enhanced, float*=t, width=\textwidth, label type=box}{box}
\usepackage[printonlyused,withpage,nohyperlinks,smaller]{acronym}
\graphicspath{{./figs/}}

\newcommand{\Strain}{\mathbf{S}^{\text{train}}}

\newcommand{\Ytrain}{\mathbf{Y}^{\text{train}}}

\renewcommand{\selectlanguage}[1]{}

\begin{document}
\title{Exoplanetary atmospheres retrieval via a quantum extreme learning machine}

\author{Marco Vetrano*}
\let\comma,
\affiliation{Universit\`a degli Studi di Palermo\comma{} Dipartimento di Fisica e Chimica - Emilio Segr\`e\comma{} via Archirafi 36\comma{} I-90123 Palermo\comma{} Italy}
\author{Tiziano Zingales}
\let\comma,
\affiliation{Dipartimento di Fisica e Astronomia “Galileo Galilei”, Università degli Studi di Padova, Vicolo dell'Osservatorio 3, 35122 Padova, Italy}
\affiliation{INAF Osservatorio Astronomico di Padova, Vicolo dell’ Osservatorio 5, 35122 Padova, Italy}
\author{G. Massimo Palma}
\affiliation{Universit\`a degli Studi di Palermo\comma{} Dipartimento di Fisica e Chimica - Emilio Segr\`e\comma{} via Archirafi 36\comma{} I-90123 Palermo\comma{} Italy}
\author{Salvatore Lorenzo}
\affiliation{Universit\`a degli Studi di Palermo\comma{} Dipartimento di Fisica e Chimica - Emilio Segr\`e\comma{} via Archirafi 36\comma{} I-90123 Palermo\comma{} Italy}

\begin{abstract}
    The study of exoplanetary atmospheres traditionally relies on forward models to analytically compute the spectrum of an exoplanet by fine-tuning numerous chemical and physical parameters. However, the high-dimensionality of parameter space often results in a significant computational overhead. In this work, we introduce a novel approach to atmospheric retrieval leveraging on quantum extreme learning machines (QELMs). QELMs are quantum machine learning techniques that employ quantum systems as a black box for processing input data. In this work, we propose a framework for extracting exoplanetary atmospheric features using QELMs, employing an intrinsically fault-tolerant strategy suitable for near-term quantum devices, and we demonstrate such fault tolerance with a direct implementation on IBM Fez. The QELM architecture we present shows the potential of quantum computing in the analysis of astrophysical datasets and may, in the near-term future, unlock new computational tools to implement fast, efficient, and more accurate models in the study of exoplanetary atmospheres.
\end{abstract}
\maketitle

\section{Introduction}


The study of exoplanetary atmospheres represents one of the most fascinating astrophysical challenges. The interpretation of atmospheric features and the analysis of absorption and emission spectra \citep{kreidberg2018global, tsiaras2018population, bruno2018comparative, mansfield2018hst, spake2018helium, sheppard2017evidence, barstow2016consistent, rocchetto2016exploring} can be performed thanks to the mathematical modeling of the exoplanet atmosphere. Generation of a forward model is crucial to simulate expected spectra from a given atmospheric composition, temperature profiles, and physical conditions. In atmospheric retrievals, forward models are compared with observed spectra to retrieve atmospheric parameters \citep{irwin2008nemesis, madhusudhan2009temperature, line2013systematic, benneke2013distinguish, cubillos2016bart, gandhi2018retrieval, lavie2017helios}. Models may be very different in their complexity, ranging from simply parameterised descriptions to complex physical simulations, usually at the price of a large increase in computing resources due to the increased precision.

Finding the right balance between computational cost and model accuracy in exoplanetary science has been a challenge. Typically, simple models enable a fast analysis at the cost of oversimplifying the atmospheric processes involved and provide biased results or miss subtle features that could be present in the data \citep{pluriel2020, pluriel2022}. More complex models can capture intricate atmospheric phenomena at a higher computational cost as they usually require advanced statistical tools to estimate atmospheric parameters \citep{feroz2008multimodal, skilling2004nested, feroz2009use, gregory2011bayesian}.

The optimization of retrieval methods and the use of machine learning techniques, which have just been introduced in several recent works, are efforts which offer a good compromise between computational cost and model accuracy. Machine Learning and deep Learning techniques have been implemented to approximate forward models at lower computational cost \citep{marquezneila2018, zingales2018exogan, cobb2019} and to detect new exoplanetary transits with several instruments \citep{pearson2018searching, shallue2018identifying, kipping2016transit}.


In the context of atmospheric retrieval, several machine learning techniques have shown the ability to make both qualitative and quantitative identifications of the spectral type \citep{waldmann2016dreaming, zingales2018exogan, marquezneila2018, zingales2018exogan, cobb2019}. These algorithms allow for fast and optimal parameter estimation, significantly improving computational time over classic retrieval methods. Furthermore, machine learning tools can be designed to return, depending on the particular problem, different output types such as posterior distributions for statistical inference, generated spectra for validation against observation, and straightforward prediction of atmospheric parameters such as temperature profiles, molecular abundance, and clouds.

One crucial aspect of reliable machine learning algorithms, are the high-quality training datasets. Such datasets require either extensive simulations or extensive observation campaigns, both options being very resource intensive. In addition to this, most machine learning models cannot make good extrapolations outside the parameter space spanned by their training data and may become inaccurate for new or complex atmospheric features. For this reason, it is important to generate robust datasets to train future algorithms, reproducing all those features that could be observed with space-based and ground-based instruments.

State-of-the-art retrieval methods are promising in exploiting low-resolution spectroscopy, given the capabilities of missions like JWST \citep{gardner2006james} and the upcoming Ariel space mission \citep{tinetti2016science}. The broad wavelength coverage and high sensitivity of JWST combined with the extensive survey of various exoplanets by Ariel yield a rich set of low-resolution spectra. In the same respect, the increased complexity of the model needed to interpret these observations underscores that the refinement of machine learning tools must go hand in hand with new challenges related to precision and scalability.

In the context of deep learning, Extreme Learning Machines (ELMs)~\cite{konkoli_reservoir_2017, huang_extreme_2004,huang_extreme_2011, wang_review_2022, markowska-kaczmar_extreme_2021} have emerged as a promising alternative to traditional approaches due to their highly efficient training protocol. 
An ELM is a neural network model in which only the final layer is trained, while all other layers, generally referred to as ``reservoir'', are randomly initialized and kept fixed. 
Despite this randomness, its inherent complexity and non-linearity enable it to extract meaningful, non-trivial features from data. 
ELMs have been successfully implemented both with randomly generated neural networks~\cite{Huang_Universal_2006} and with a variety of physical systems~\cite{soriano_delay-based_2015, bhovad_physical_2021, nakajima_information_2015, coulombe_computing_2017, goudarzi_dna_2013, tanaka_recent_2019, nokkala_high-performance_2022}.

With the advent of Quantum Machine Learning (QML)~\cite{biamonte2017quantum, schuld2015introduction, cerezo2022challenges, Banchi_Generalization_2021} ELMs have been extended to quantum reservoirs, leading to the development of Quantum Extreme Learning Machines (QELMs). 
Such models have demonstrated their effectiveness as a powerful tool for processing both quantum and classical data~\cite{vetrano2024state, mujal_opportunities_2021, suprano_experimental_2023, govia_quantum_2021, martinez-pena_quantum_2023, ghosh_quantum_2019,ghosh_quantum_2019-1, martinez-pena_dynamical_2021, ghosh_realising_2021, s_ghosh_reconstructing_2021, mujal_time-series_2023, krisnanda_creating_2021,garcia-beni_scalable_2023,martinez-pena_information_2020},  outperforming their classical counterpart in terms of computational resources ~\cite{mujal_opportunities_2021, fujii_harnessing_2017, xiong_fundamental_2023, nakajima2019boosting}. 
However, processing high-dimensional data with QML remains a challenge from both the simulation and the experimental point of view. In fact, while classical simulations of many-body systems are very expensive in terms of computational resources, real quantum devices still suffer from decoherence noise caused by unwanted interactions with the environment~\cite{preskill2018quantum, lau2022nisq}. 
In some cases, the challenges posed by high-dimensionality can be addressed by patching data arrays to reduce the dimensions of the task, effectively breaking it into subtasks~\cite{huang2021experimental, tsang2023hybrid}.

Inspired by the architecture used in~\cite{huang2021experimental}, in this work, we construct a QELM to extract information from synthetic spectra generated by TauREx~\cite{alrefaie2021,waldmann2015tau} 
with the key focus to minimize the quantum resources needed to ensure the  fault-tolerance with near-term implementations. 
To test QELM's robustness, the algorithm is evaluated on datasets within the JWST spectral range, both with and without added artificial shot noise, to assess its response to realistic data imperfections. Additionally, its fault tolerance is demonstrated by implementing the JWST dataset on the IBM Fez quantum computer, validating its performance on real quantum hardware and its potential for practical applications.

The article is organized as follows: in~\cref{sec:QELM} we briefly review ELMs and QELMs; in~\cref{sec:methods} we give a detailed explanation of the structure of the datasets employed in this work, its pre-processing and the implementation of our QELM framework; in~\cref{sec:results} we show the results obtained by running the algorithm on IBM Fez and show the fault tolerance of the algorithm.
Finally in~\cref{sec:Conclusions} we derive our conclusions.


\section{QELMs in a nutshell}
\label{sec:QELM}
{\it ELM - }An ELM is a fast, efficient machine learning algorithm~\cite{goodfellow_deep_2016} that uses single-hidden-layer feedforward neural networks with randomly assigned weights and biases, called ``{\bf{reservoir}}", requiring only the output weights to be learned through a final linear post-processing step.

The reservoir can be described as a complex function $\mathcal{R}:\mathbb{R}^{D_{\rm in}}\to\mathbb{R}^{D_{\rm out}}$ mapping the input data in a higher dimensional space. Thus the complete function applied by an ELM can be written as the composition $f^{\text{ELM}}_{\bf W}{=}\mathbf W\circ \mathcal{R}$, where $\mathbf{W}$ is the trained linear map applied by the output layer. 

Given the fixed and untrained nature of the reservoir function $\mathcal{R}$, ELMs can be implemented using recurrent neural networks with random initialized weights~\cite{lukosevicius_reservoir_2009} or with physical hardware~\cite{soriano_delay-based_2015, bhovad_physical_2021, nakajima_information_2015, coulombe_computing_2017, goudarzi_dna_2013, tanaka_recent_2019, nokkala_high-performance_2022}. 
This approach simplifies the learning process to solving a linear regression problem, eliminating the need for classical optimization algorithms to fine-tune the entire network's parameters. As a result, it achieves generalization much more efficiently.

ELMs belong to the class of supervised learning models. Given a training dataset $\{\Strain,\Ytrain\}$, the training process consists in finding the best $\mathbf W$ which approximates
\begin{equation}\label{eq:linear_problem_training}
   f^{\text{ELM}}_{\bf W} (\Strain)\equiv\mathbf W \mathcal{R}(\Strain) =\mathbf Y^{\rm train},
\end{equation}
where $\Strain \equiv \{s^{\rm train}_k\}$ is the ensemble of training input data, $\mathcal{R}(\Strain)$ is the matrix whose $kth$-column corresponds to the outcomes of the reservoir relative to the input $s^{\rm train}_k$ and $\mathbf Y^{\rm train}$ is the matrix containing the target associated to each input data. More specifically, $\mathcal{R}(\Strain)$ results to be a $D_{\rm out}\times D_{\rm train}$ matrix, while $\mathbf Y^{\rm train}$ is $D_{\rm feat} \times D_{\rm train}$.
We indicate with $D_{\rm out}$ the number of reservoir outcomes, with $D_{\rm train}$ the number of training samples and with $D_{\rm feat}$ the number of target features.
Having in mind an implementation  with a physical reservoir, the input data $\Strain$ can be encoded in a perturbation of the reservoir state, and the response of the reservoir is described by $\mathcal{R}(\Strain)$.

The training process involves solving the linear system in~\cref{eq:linear_problem_training}. A common technique  is to use the the Moore-Penrose pseudo-inverse:
\begin{equation}\label{eq:classical_training}
\mathbf{W}=\Ytrain \mathcal{R}(\Strain)^+,
\end{equation}
where $\mathcal{R}(\Strain)^+$ is , computed via the Singular Value Decomposition (SVD) $\mathcal{R}(\Strain)^+ = U\Sigma^+ V^\dagger$, with $U, V$ isometries and $\Sigma^+$ a positive diagonal matrix containing the inverse of the singular values of $\mathcal{R}(\Strain)$~\cite{serre2010matrices}.

\begin{figure}[h!]
    \centering
    \includegraphics[width=\linewidth]{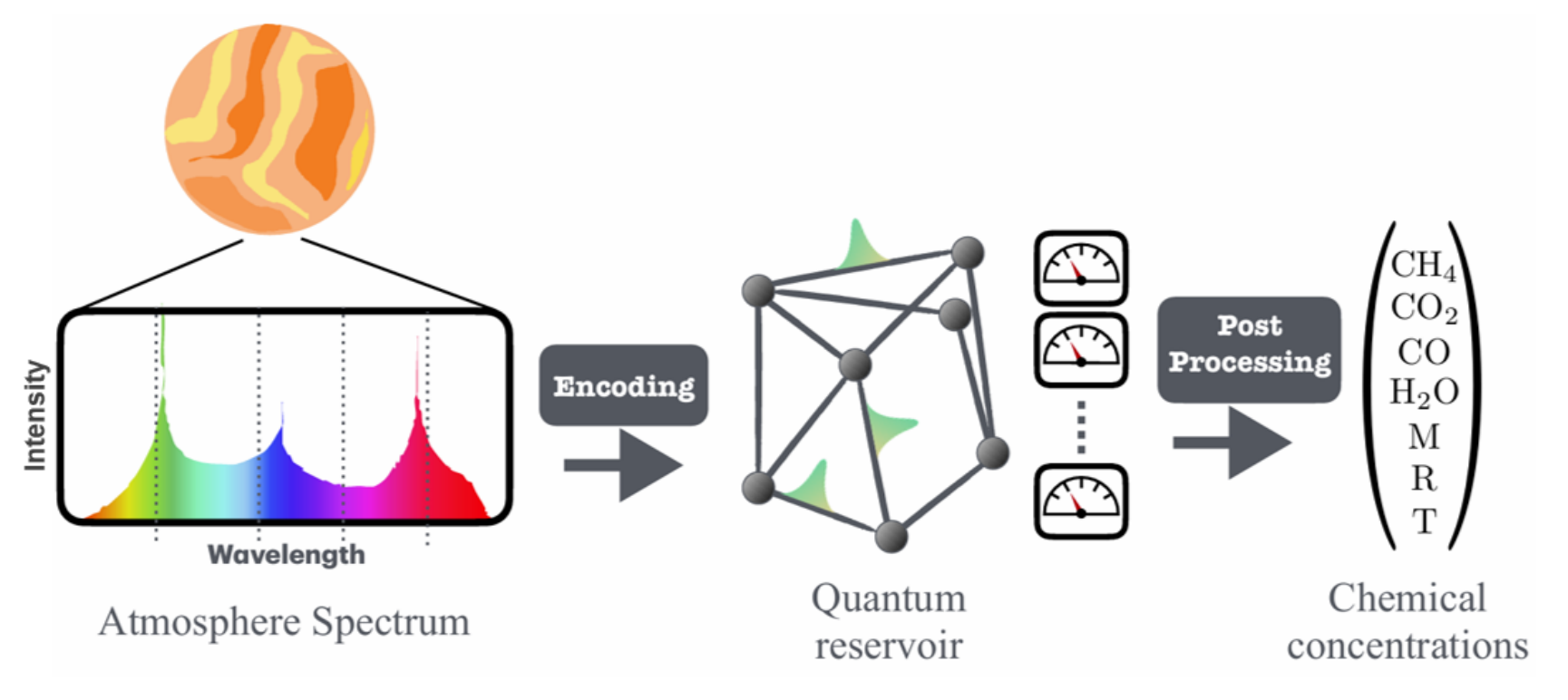}
    \caption{Pictorial representation of how QELM on classical data works. Classical data are encoded in a physical system from which we collect a set of measurements. The outcomes undergo a linear post-processing which maps the outcomes of the reservoir to the output of task.}
    \label{fig:QELM}
\end{figure}

{\it QELM - }The transition from classical to quantum extreme learning machines is accomplished by replacing the classical reservoir with a quantum system. 
In this framework, the reservoir function $\mathcal{R}$ is replaced by a quantum dynamical map $\Phi$ followed by a measurement step~\cite{vetrano2024state,innocenti_potential_2023, mujal_opportunities_2021}.
Furthermore, whenever the input data are not purely quantum, they need to be encoded in a quantum state in order to be processed by the reservoir~\cite{mujal_opportunities_2021, de2024harnessing, mujal_time-series_2023}, giving us a dataset of training states ($s^{\rm train}_k\rightarrow\rho^{\rm train}_k$). 
If we assume that the measurement step is performed evaluating the expectation values of a Positive Operator Valued Measure (POVM) ${\Pi_i}$, we have the following correspondence:
\begin{equation}\label{eq:quantum_case}
    \mathcal{R}(\Strain) \rightarrow P^{\rm train}_{ik} = \Tr{\Pi_i\Phi[\rho^{\rm train}_k]}
\end{equation}
where ${\mathbf{P}}^{\rm train}$ is the probability matrix relative to the training states. The dimension of $\mathbf{P}$ entirely depends on the number of measured operators on the reservoir and on the dimension of the training dataset, i.e $D_{\rm out}\times D_{\rm train}$. It is important to stress the fact that, since both the quantum map $\Phi$ and the post-processing layer are linear in the input quantum states, QELMs {\it cannot} be affected by over-fitting. On the other hand, being the action of classical reservoir not necessarily linear on the input data, this is not generally true for classical ELMs. A pictorial representation of QELM is shown in~\cref{fig:QELM}.

Finally, to evaluate the system's generalization capabilities, we use a testing dataset along with its associated probability matrix $\mathbf{P}^{\rm test}$. The performance is assessed using a task-specific metric; for this work, we employ the relative error between the reconstructed and testing features:

\begin{equation}\label{eq:metric}
    \epsilon = \frac{(y_k^{test}-y^{pred}_k)^2}{(y_k^{test})^2}*100.
\end{equation}

\section{Methodology} \label{sec:methods}
\begin{figure*}
    \centering
    \includegraphics[width=0.8\linewidth]{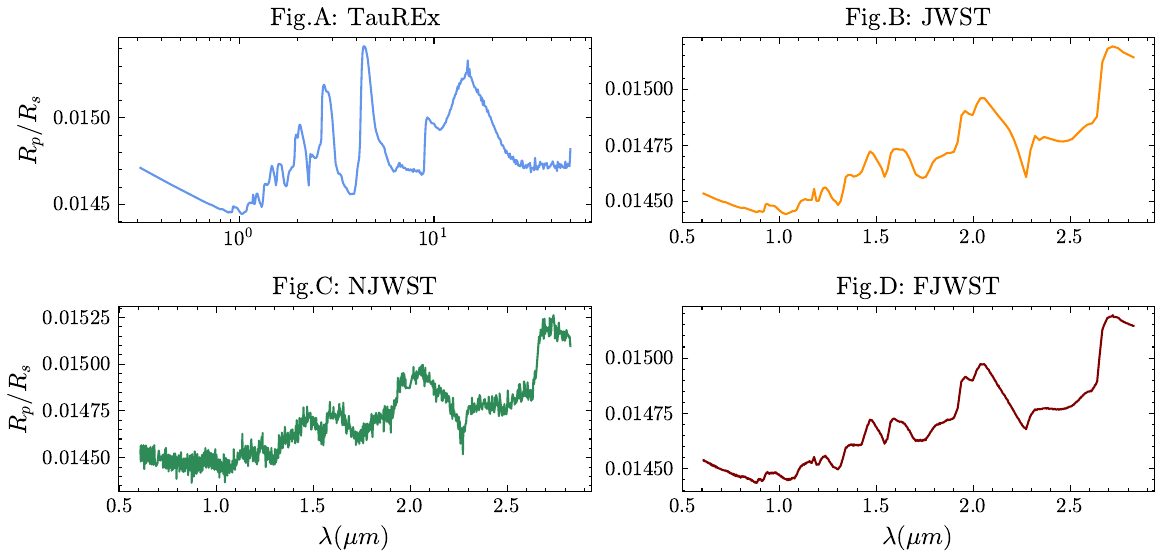}
    \caption{Example of a spectrum produced by TauREx ({\bf Fig.A}), interpolated on JWST spectral range ({\bf Fig.B}), with the addition of shot noise ({\bf Fig.C}) and with the addition of a principal component filter taking 10 components in account ({\bf Fig.D}). As can be seen in the spectra, adding the filter the presence of noise still changes a bit the spectrum with respect to its clean counterpart in {\bf Fig.B}}
    \label{fig:spectra}
\end{figure*}
In this section, we review the structure of the training dataset, its pre-processing and the working of the QELM.

\subsection{Dataset}
As mentioned the aim is to build a quantum extreme learning framework able to make atmospheric retrieval of exoplanets. 
In particular, we are going to process datasets of spectra generated by TauREx~\cite{alrefaie2021,waldmann2015tau}. TauREx is an open-source Bayesian code for exoplanetary atmospheric modeling and retrieval whichhandle allows for the generation and retrieval of transmission, emission and reflection spectra over a wide range of compositions and physical parameters. TauREx computes radiative transfer on exoplanetary atmospheres taking into account different chemical and physical conditions to produce the resulting spectrum. Its object-oriented structure allows the inclusion of many different atmospheric scenarios and it is widely used in the exoplanetary community.

The dataset we used in this context is the same used in \citep{zingales2018exogan}.
It depends in particular on 4 parameters for the molecular volume mixing ratios ($CH_4, CO_2, CO, H_2O$) and 3 physical parameters $M, R, T$, respectively, planetary mass, radius and equilibrium temperature. 
We assume a constant vertical profile for each volume mixing ratio and isothermal temperature-pressure profile.
Each parameter has its lower and upper bounds and the intervals are divided in 10 discrete values. 
Each spectrum is generated by TauREx in the spectral range $[0.3, 50]\mu m$, divided in 515 spectral bins, considering a combination of randomly selected values for each parameter. 
Thus we have $10^7$ spectra available for training and testing. 
In this work we process a sampled subset of $10^4$ spectra for both the training and the
testing stage. 
All the details regarding the lower and upper bound for each of the atmospheric parameters are summarized in~\cref{tab:parameters}. This dataset will be labeled as {\bf TauREx}.

In order to test QELM in a more realistic scenario, we considered a spectrum of HAT-P-18b~\cite{fu2022water}, measured by JWST and processed the {\bf TauREx} spectra by interpolating the same spectral binning of JWST in the spectral range $[0.6,2.8]\mu m$. This second dataset will be labeled as {\bf JWST}.

The {\bf JWST} spectra have also been processed considering shot noise obtained by a source of photons at $T_\star {=} 6460K$. The shot noise has been computed as shown in ~\cite{cowan2015characterizing}. This last dataset will be labeled as Noisy JWST ({\bf NJWST}).

Finally, we considered one last dataset by filtering out the shot noise with a pre-processing shown in the next subsection and we are going to label it as Filtered JWST ({\bf FJWST}). In~\cref{fig:spectra} we show an example of all the previously described data.

\begin{table}[h!]
    \centering
    \begin{tabular}{|c|c|c|}
        \hline
        Variable &  Lower Bound & Upper Bound\\
        \hline
         $CH_4$  &     $10^{-8}$  &   $10^{-1}$  \\
         \hline
         $CO_2$  &     $10^{-8}$  &   $10^{-1}$  \\
         \hline
         $CO$    &     $10^{-8}$  &   $10^{-1}$  \\
         \hline
         $H_2O$  &     $10^{-8}$  &   $10^{-1}$  \\
         \hline
         $M$     &     $0.8M_J$ &   $2.0M_J$ \\
         \hline
         $R$     &     $0.8R_J$ &   $1.5R_J$ \\
         \hline
         $T$     &     $1000K$  &   $2000K$  \\
         \hline
    \end{tabular}
    \caption{The spectra considered in this work have been produced considering atmospheric parameters in the upper intervals. Each interval has been divided in 10  values. The spectra are produced considering all the $10^7$ possible combination of these values.}
    \label{tab:parameters}
\end{table}
\subsection{Data pre-processing}

The first problem to solve in the context of quantum machine learning in general is the encoding of high dimensional data in a Hilbert space which is possibly small. The choice of  encoding will impact on the quantum resources needed  to solve a given task. Furthermore in the context of Near Intermediate Scale Quantum (NISQ) devices, solving a task with too many quantum resources, for instance deep quantum circuits, results in very noisy outputs which do not allow for the implementation on real near-term quantum devices. In our context, by ``quantum resources'' we mean the number of qubits and the number of quantum operations performed on a quantum system to complete a task. 
Therefore, when handling high dimensional data, the common approach consists in reducing their dimension with either convolutional neural networks or principal component analysis (PCA)~\cite{de2024harnessing, huang2021power, rebentrost2014quantum,senokosov2024quantum} to extract the most relevant and discernible features of a dataset. For instance, PCA is a linear transformation which consists in a change of coordinates, with the aim to find the frame in which the variance of the data is maximized.

On the other hand, in the context of generative quantum machine learning such as Quantum Generative Adversarial Networks (qGANs)~\cite{huang2021experimental}, the number of qubits determines also the resolution (number of pixels) of the images which the qGAN is able to produce. Thus in this context it is not possible to minimize the number of qubits; nonetheless, it is still possible to limit the depth of the quantum circuits (number of gates) by factoring the task into a set of sub-tasks. This reduces the long range interactions between qubits and thus the number of quantum operations applied in a circuit.

To handle the spectra, in this work we decided to use a hybrid approach which combines both the dimensionality reduction and the organization of the task into sub-tasks. In particular, we divided the spectral range according to the major water bands in the IR while also taking into account the instrument pass-bands of JWST/NIRISS, NIRcam, MIRI and Hubble/WFC3, for a total of $N_{\text{\bf TauREx}} {=} 14$ spectral bands, which from now on we are going to address as ``\textit{patches}". The number of patches has been reduced to $N_{\text{\bf JWST}} {=} 8$ when considering only the JWST/NIRISS spectral range. Beyond the need to factorize the problem, this operation has also been performed in order to normalize the patches one by one and amplify their spectral features. 

In the {\bf JWST} case, we also test the model considering the artificial noise added to the dataset. The shot noise is computed as in~\cite{cowan2015characterizing}. 
We assumed that measurement noise is dominated by quantum detection noise following Poisson statistics with equal mean and variance between $\sim$0.6 and 2.8~$\mu$m. We estimated the mean number of collected photoelectrons to be:
\begin{equation}
N_\mathrm{ph} = \frac{\pi \tau \Delta t}{hc} \left( \frac{R_\star D}{2d}\right)^2 \int_{\lambda_1}^{\lambda_2} B(\lambda, T_\star)\lambda d \lambda,
\label{photon_noise}
\end{equation}
where $\lambda_1$ and $\lambda_2$ are the limiting wavelengths of the spectral bin, $d$ is the distance of the star (we took here 270 pc by default), $R_\star = 1.46 R_{{\text Sun}}$, $T_\star = 6460K$, $d {=} 270{\text pc}$ and $\Delta t = 21340s$, respectively. The parameters $D = 16$, $\tau = 0.4$, and $\Delta t$ are the diameter of the telescope, the performance of the system, and the integration time, respectively, whose values were fixed for JWST according to \citet{cowan2015characterizing}. 
Since systematics may prevent us from reaching a 10 ppm precision with the JWST, we assumed a floor noise of 30 ppm throughout the whole spectral domain with a normal distribution \citep{greene2016} from 0.6 to $\sim 2.8 \, \mu$m, wherever the noise was lower than 30 ppm.

In this last case, before proceeding with the extraction of the principal components, the spectra are filtered using a PCA considering the first 10 components to reconstruct the filtered data. The number of components used to filter out the noise was chosen by examining the cumulative explained variance of the principal components, which gives an idea on how much information we are capturing by retaining a certain number of features. It is important to stress the fact that the filtering procedure in this case is necessary since applying the pre-processing as we do in the other two cases amplifies both the spectral features and the noise, leading to worse reconstructions.

The last step of the pre-processing is to extract the $M$ principal components of each of the spectral bands in order to reduce the dataset dimension. The patch division will give us the opportunity to implement, as we will show in the next subsection, the QELM in a fault tolerant way; on the other hand, applying the PCA to the normalized patches allows us to extract features which are comparable in magnitude.
Thus, from each patch we extract a M-dimensional vector $\mathbf{x^i}$ of principal components, ending up with $N_p$ with $N_p\in\{N_{\text{\bf TauREx}}, N_{\text{\bf JWST}}\}$ to which we also add another vector containing the global maximum and minimum of each spectrum and its average value. All these vectors are joined in a matrix $\mathbf{X}$
\begin{equation}\label{inputX}
    \mathbf{X} = \begin{bmatrix}
        \mathbf{x}^1 \\
        \mathbf{x}^2 \\
        ...\\
        \mathbf{x}^{N_p+1}
    \end{bmatrix}
\end{equation}

\subsection{Reservoir dynamics}

In order to limit the number of qubits and the depth of the reservoir, with a similar reasoning of~\cite{huang2021experimental, tsang2023hybrid} we consider a factorized reservoir composed in the following way: the first sub-reservoir encodes the average of the spectra and their global maxima and minima; on the other hand, the other $N_p$, with $N_p \in \{N_{\text{\bf{TauREx}}} = 14, N_{\text{\bf{JWST}}} = 8\}$ have the objective to process the incoming information from each patch of the spectra.

In order to process a $M$-dimensional dataset, we will need to initialize a quantum register of $Q\geq M$ in the $|{\bf 0}\rangle$ state and build a quantum circuit with the following structure

\begin{itemize}
    \item {\bf encoding layer}: the input data are mapped into a set of angles $\theta \in [0,2\pi]$ and encoded in a $R_X$ rotation acting on the first $M$ qubits;
    \item {\bf reservoir layers}: a simple multilayered reservoir composed of randomized rotations RY on each qubit, a layer of CNOT gates and a third layer of random RY rotations;
    \item {\bf measurement layer}: collection of the measurements of all the qubits, which gives the outcome probabilities of measuring all the possible $2^Q$ states.
\end{itemize}

\begin{figure}[h!]
    \centering
    \includegraphics[width=\linewidth]{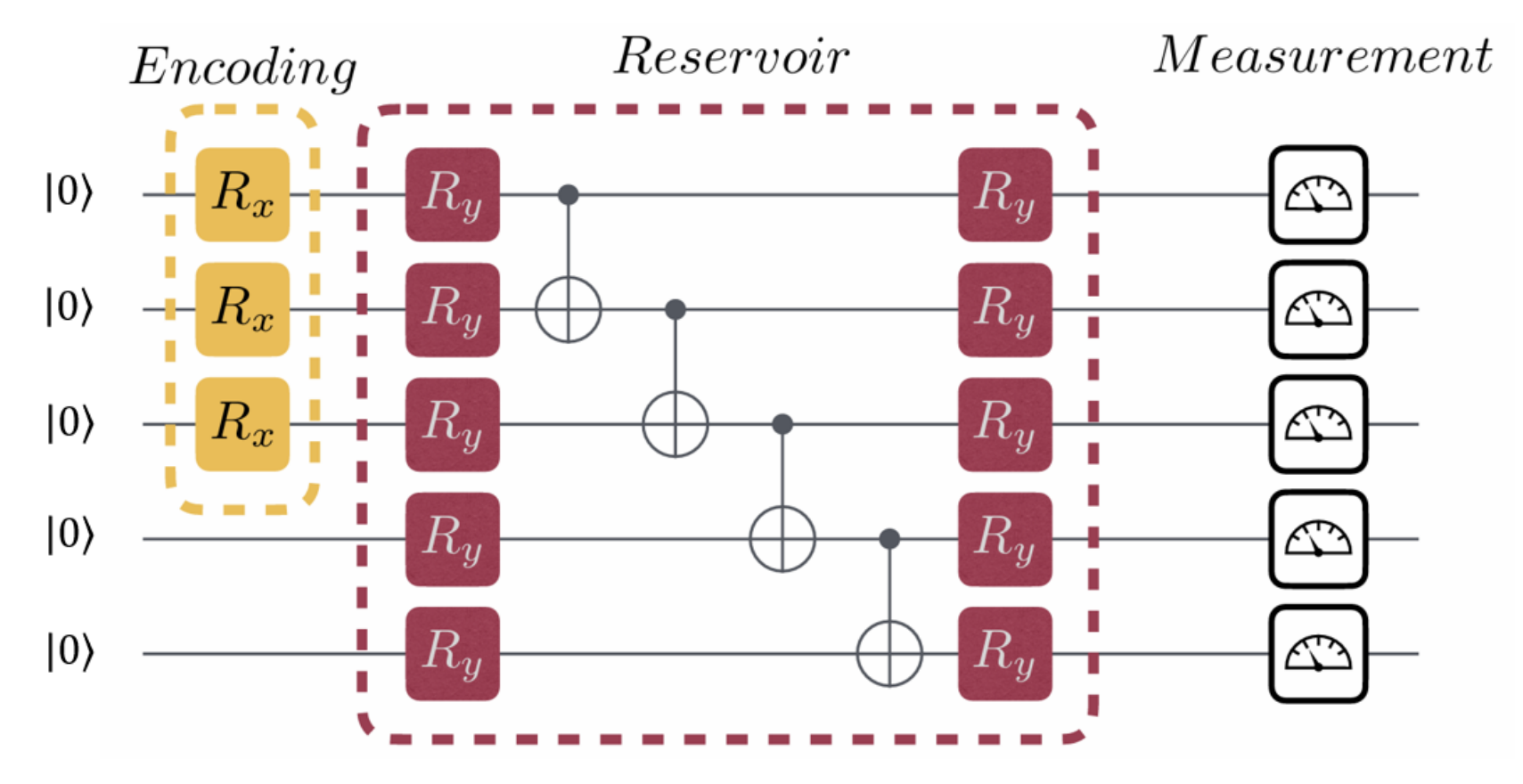}
    \caption{Pictorial representation of a single reservoir. The {\bf encoding} of the classical data is performed by means of RX gates whose rotation angles depend on the principal component extracted by each patch; the {\bf reservoir layer} consists of random $R_Y$ rotations, a set of CNOT and other random $R_Y$ rotations. Finally the outcome probabilities are collected by measuring the final state.}
    \label{fig:reservoir}
\end{figure}

In~\cref{fig:reservoir} we show a pictorial representation of the action of the encoding gates and a reservoir.

{\it Encoding layer -}  Let $\mathbf{O}$ be a single qubit operator, from now on we will denote with $O_k$ the tensor product:

\begin{equation}
O_k \equiv \mathbb{I}\otimes\mathbb{I}\otimes \hdots \otimes \overbrace{\mathbf{O}}^{k\text{-th}} \otimes\hdots \otimes\mathbb{I}\otimes\mathbb{I}
\end{equation}

The $R_X$ gate performs local phase rotation on a qubit along the X axis, thus RX is
\begin{align}
    R_X(\theta) = e^{-i\theta \sigma^x_k} 
\end{align}
Therefore, the encoding layer applies a unitary evolution $U^i_{Enc} =\prod^M_{k = 1}R_X(x^i_k)$, where we have encoded the k-th principal components of the i-th patch in the rotation angle of RX.
\begin{figure*}
    \centering
    \includegraphics[width=0.95\linewidth]{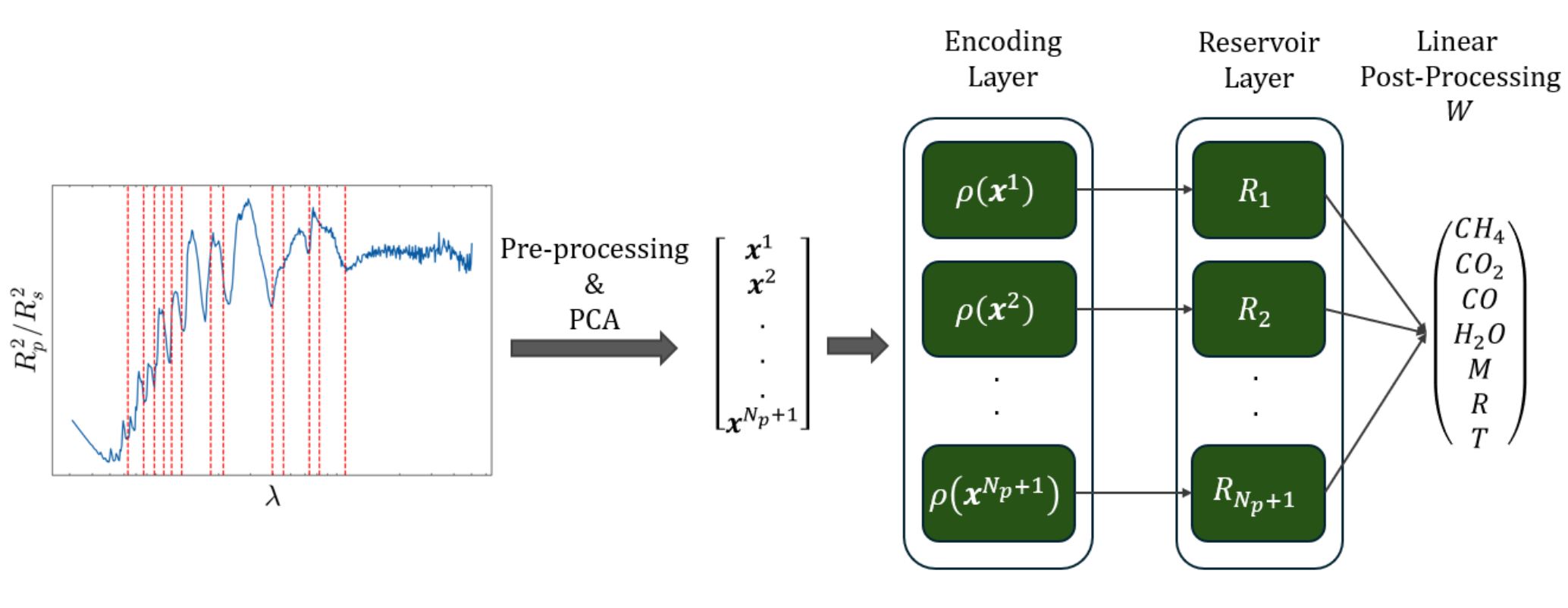}
    \caption{Resuming scheme of the whole processing: spectra are patched according to~\cite{zingales2018exogan} and pre-processed by the processing layer consisting of an interpolator (for JWST spectra), a normalization layer and a principal component analysis layer acting on each of the patches. The components are encoded in a reservoir as shown in~\cref{fig:reservoir} and the outcome probabilities are than post-processed via the output layer to obtain the atmospheric parameters.}
    \label{fig:final scheme}
\end{figure*}

{\it Reservoir layer - } While in the encoding layer we only acted on a subset of qubits, in the reservoir we work with all of them. In particular, we perform random local rotations along the Y axis by applying the RY gate on each qubit, with RY
\begin{align}
    RY(\theta) = e^{-i\theta \sigma^y_k} 
\end{align}
The CNOT gates in the reservoir defined as 
\begin{equation}
    {\text CNOT^{k+1}_k} = |0\rangle\langle0|_k+|1\rangle\langle1|_k  \sigma^x_{k+1}
\end{equation}
act on consecutive pairs of qubits $k, k+1$ with a conditional dynamics where the first qubit of the pair acts as the control and the second qubit as the target.

Then we apply a second layer of random RY rotations. Thus, each reservoir in the reservoir layer consists of a unitary evolution $U^i_{Res} {=} \prod^Q_{k = 1} RY(\alpha_k)\prod^{Q-1}_{k=1}{\text CNOT}^{k+1}_k\prod^Q_{k=1}RY(\beta_{k})$.

{\it Measurement Layer -} In the last layer we perform a projective measurement on each individual qubit, obtaining a set of outcome probabilities $\mathbf{p}^i$ for each reservoir:

\begin{equation}\label{eq: outcome}
    \mathbf{p}^i \equiv\{p^i_m\} =\langle\mathbf{0}|U^{i\dag}_{Enc}U^{i\dag}_{Res} \,\,{\Pi_m}\,\,U^i_{Res}U^i_{Enc}|\mathbf{0}\rangle
\end{equation}
The output from each reservoir $\mathbf{p}^i$, corresponding to the processed outcome of the respective patch $\mathbf{x}^i$ in \cref{inputX}, are aggregated into a single output matrix $\mathbf{P}$ structured as:
\begin{equation}\label{eq:outcomeprobabilities}
\centering
    \mathbf{P} = \begin{bmatrix}
    \mathbf{p}^1 \\
    \mathbf{p}^2 \\
    ...\\
    \mathbf{p}^{N_p+1}
    \end{bmatrix}
\end{equation}
The aggregated output vector $\mathbf{P}$ is then processed by a linear output layer, which extracts the atmospheric parameters of the spectra. A pictorial representation of the QELM model used in this work is visualized in~\cref{fig:final scheme}.

We consider successful retrievals those with relative error (\cref{eq:metric}) $\epsilon\leq5\%$. Thus we define the accuracy $A$ as:
\begin{equation}
    A = \frac{1}{D_{\text {test}}}\sum^{D_{\text{test}}}_i\delta(\epsilon_i)
\end{equation}
\[
\delta(\epsilon_i) =
\begin{cases}
1, & \text{if } \epsilon_i \leq 5\%, \\
0, & \text{otherwise}.
\end{cases}
\]

\section{Results}\label{sec:results}

We tested our algorithm on {\bf IBM Fez}, a quantum computing device with 156 qubits. In particular, we used this device to process {\bf JWST} data, which we argue to be a good compromise between the more complete data generated by TauREx (\cref{fig:spectra}{\bf a}) and the more realistic {\bf NJWST} data (\cref{fig:spectra}{\bf c}). The reservoir layer consists of $N_{\text{\bf JWST}} + 1  {=} 8+1$ reservoirs, processing in parallel the principal components obtained by the $8$ patches, plus $1$ for the average of the spectra, global maxima and minima.
From each patch we extracted $M {=} 5$ principal components and worked with a dataset of dimension $D {=} 4080$, with the $75\%$ used for training while the remaining $25\%$ for testing. 
\begin{figure}
    \centering
    \includegraphics[width=\linewidth]{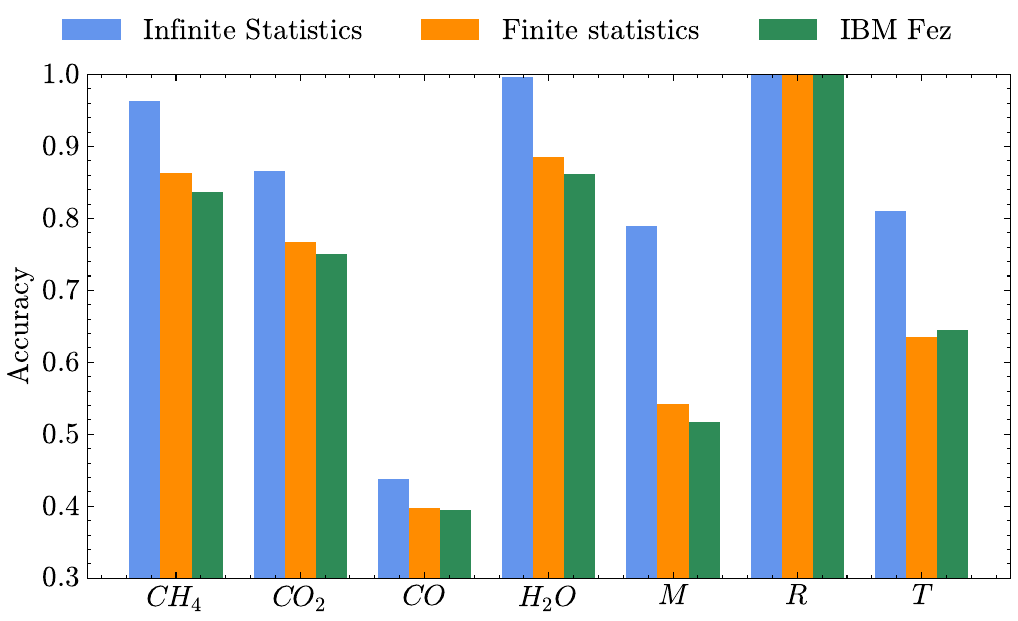}
    \caption{Results obtained by processing the {\bf JWST} dataset using IBM Fez. The results have been obtained using reservoirs of 5 qubits each and $M {=} 5$. In this context we considered a dataset of $D {=} 4080$ spectra, with the $75\%$ used for training and the rest for testing. Both the results obtained on hardware and the finite statistics simulation were obtained with 20000 shots.}
    \label{fig:hardware results}
\end{figure}
As we show in~\cref{App:Training and Feature Test}, altough these are not the most ideal condition to fully train the output layer and to capture all of the information needed from the spectra, they  nonetheless are enough to both show fault tolerance and provide good results.

In order to parallelize the action of the $9$ reservoirs, we designated $9$ subsets of 5 qubits on the QPU, each used to process the incoming data from a patch of the spectra or the global properties as visualized in~\cref{fig:final scheme}. Each qubit in the reservoirs processing the principal components was encoded with one as shown in~\cref{fig:reservoir}.
After the processing we end up with a vector of outcome probabilities of the qubits of the whole quantum processor, but of course in this context we are only interested in retaining the outcome probabilities of the qubits of each reservoir. From each reservoir we then obtain a $2^M = 32$- dimensional vector $\mathbf{p_i}$ which are then all joined together in a single $(N_{\text{\bf JWST}}+1)*2^M$-dimensional vector (\cref{eq:outcomeprobabilities}) $\mathbf{P}$ which undergoes the usual linear post-processing procedure shown in~\cref{eq:linear_problem_training}. The processing of the whole dataset took about $6$ hours of computational time.

In~\cref{fig:hardware results} are visualized the accuracies in the retrievals performed with the QELM implemented with IBM Fez. The blue bars represent the result obtained with an infinite statistics simulation, meaning that the processing is purely mathematical, without taking any sampling statistics into account when measuring the outcome probabilities; the orange bars represent the results obtained by simulating a sampling statistics of $20000$ shots, meaning that each data has been processed $20000$ times by the reservoir layer and thus we have the same number of measurement outcomes from the measurement of the state of the qubits, from which we extract the outcome probabilities; at last, the green bars represent the result obtained employing IBM Fez to process the dataset with the same sampling statistics of the simulation. The numerical results of the accuracy~\cref{tab:simulation_results} show that with the current dataset and principal components, we are be able to retrieve the radius of all the exoplanets of the dataset within a relative error of $5\%$ and the concentration of $CH_4$ and $H_2O$ of almost all the dataset within the same threshold. Furthermore, we are able to reconstruct also $CO_2$ for the $86\%$ of the dataset and the other following with decreasing accuracy. 

\begin{table}[h!]
\centering
\begin{tabular}{|c|c|c|c|c|c|c|c|}
\hline
 Accuracy (\%)& $CH_4$ & $CO_2$ & $CO$ & $H_2O$ & $M$ & $R$ & $T$ \\ \hline
\text{Infinite Stat} & 96.3 & 86.6 & 43.8 & 99.7 & 78.9 & 100.0 & 81.0 \\ \hline
\text{Finite Stat} & 86.4 & 76.7 & 39.7 & 88.5 & 54.2 & 100 & 63.5 \\ \hline
\text{IBM Fez}    & 83.6 & 75.1 & 39.5 & 86.2 & 51.7 & 100.0 & 64.5 \\ \hline
\end{tabular}
\caption{Comparison of simulation and IBM Fez accuracy, visualized in~\cref{fig:hardware results}, in retrieving the atmospheric parameters within a relative error $\epsilon \leq 5\%$.}
\label{tab:simulation_results}
\end{table}

In~\cref{fig:hardware results} one can be observed how, at finite statistics, we lose about the $10\%$ of accuracy in  all of the parameters with the highest retrieving accuracy, which is still acceptable and can be improved with a higher number of shots. What is truly remarkable is that there is almost no difference between the results obtained with a finite finite statistics simulation and those obtained with IBM Fez, proving successfully the fault tolerance of the algorithm.
In~\cref{App:Training and Feature Test} and~\cref{App: Simulations Results} we show that all of the results can be improved considering a larger training dataset and a larger number of principal components.

\begin{figure*}[ht!]
    \centering
    \includegraphics[width=\linewidth]{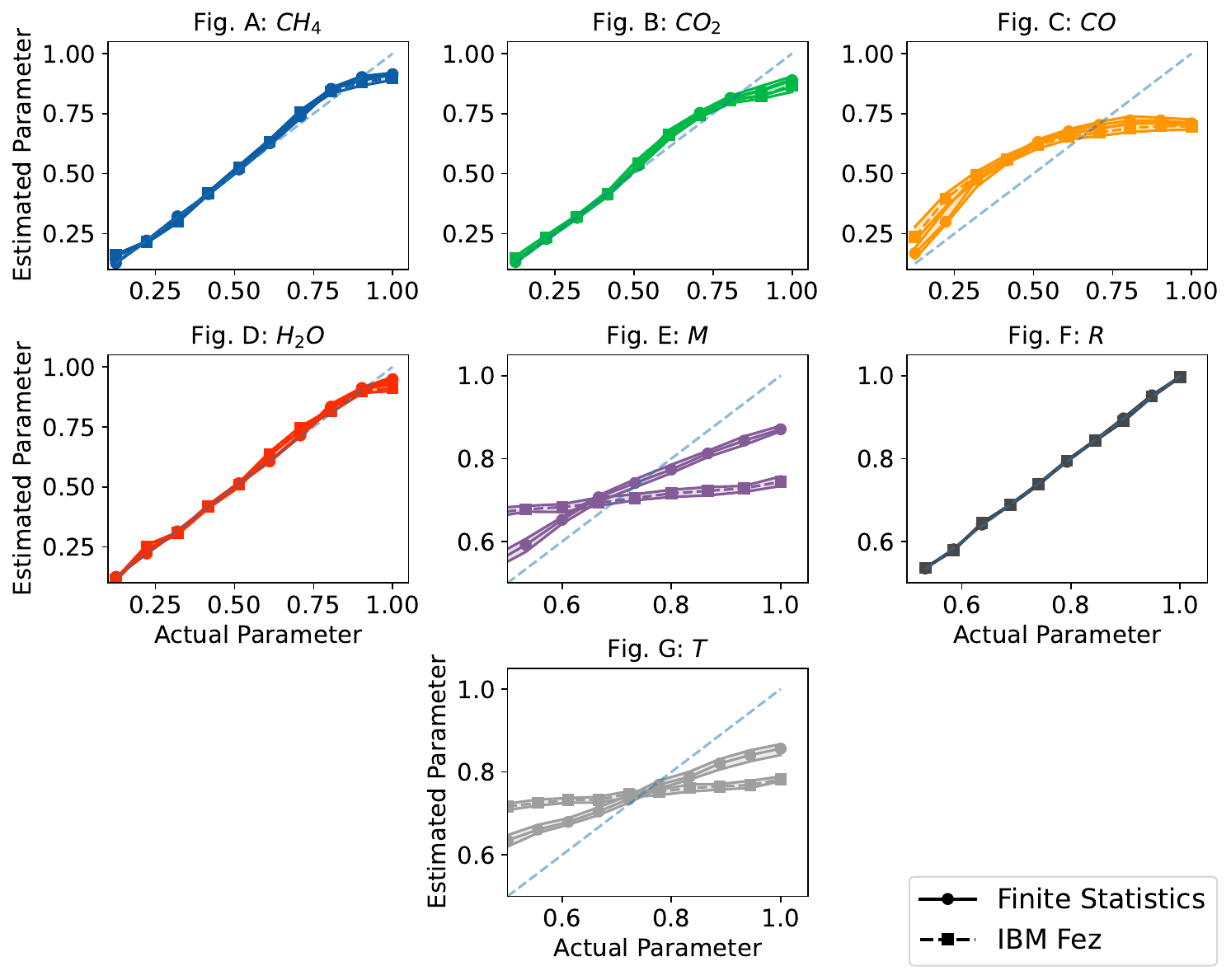}
    \caption{Comparison between the parameters estimated with QELM and the real ones. The spectral data are generated using atmospheric parameters which have a discrete distribution in the intervals reported in~\cref{tab:parameters}. The estimation of each value has been obtained by making a bootstrap of the distributions obtained for each value. In the plot we show the lower and upper bounds of the confidence interval obtained with the bootstrap, with a confidence level of 95\%. The solid lines marked with dots represent the estimation obtained with a finite statistics simulation using the {\bf JWST} dataset with $M {=} 5$ and a dataset of $D {=} 4080$ spectra divided in $75-25\%$ among training and testing. The dashed lines marked with squares represent the processing of the same dataset made with IBM Fez.}
    \label{fig:parameter_estimation}
\end{figure*}

As previously discussed, each spectrum has been generated by TauREx considering the input atmospheric parameters distributed accordingly to a discrete distribution comprising ten possible values for each parameter. To evaluate how effectively the QELM estimates these ten categories, we employed a bootstrap method to resample the unknown underlying distribution of the estimated parameters of the training set. This gave us an estimate of how well the QELM was able to discern  spectra with different atmospheric parameters and associate an error to its predition. The results are visualized in~\cref{fig:parameter_estimation}: each panel shows the median parameters obtained by re-sampling the distribution of the outcome parameters estimated with the QELM compared with the actual ones. 
The error bar corresponds to the confidence interval of the bootstrapped distribution, which covers $95\%$ of the distribution. The setup in terms of utilized dataset and extracted principal components is the same as~\cref{fig:hardware results}. From the figure, we can better appreciate how both the QELM trained with the simulated and IBM Fez processed data is able to perfectly discern spectral data from exoplanets with different radius and almost perfectly estimating the concentration of all chemical compounds in its atmosphere, except for $CO$. We however  observe that even though our QELM is not able to exactly reconstruct the concentration of $CO$, it is still able to discern exoplanets with low concentration from the others, with decreasing accuracy as the $CO$ increases. On the other hand, the QELM fails to reconstruct mass and temperature, estimating about the same value for all of the spectra. The results on this side are better for the finite statistics simulation, but still the QELM is not fully capable even in this case to extrapolate the information about these physical parameters. In~\cref{App: Simulations Results} we are going to show from the simulation results that this issue could be solved with higher statistics. It is important to stress that the results obtained with IBM Fez are consistent with the simulation result, thus proving fault tolerance to decoherence noise.

\section{Conclusion and future works}\label{sec:Conclusions}

In this work, we have shown that QELM is a promising technique for the retrieval of exoplanetary atmospheres, with both good accuracy and fast learning. Furthermore, we have shown the fault tolerance of the algorithm on an actual current quantum device, and we are confident that with the advance of quantum technologies we are going to achieve even faster data processing with fewer simplifications and higher accuracy. Furthermore, we are confident that with the current technology we can already exploit this algorithm to integrate the current forward models using the QELM solutions as ansatz for classical atmospheric retrieval techniques, leading them to converge faster. The QELM architecture presented in this work highlights the potential of quantum computation in the analysis of astrophysics datasets and could unlock, in the near future, the ability to perform efficient spectral retrieval using more complex atmospheric models.

\section*{Acknowledgments}
TZi acknowledges support from CHEOPS ASI-INAF agreement n. 2019-29-HH.0, NVIDIA Academic Hardware Grant Program for the use of the Titan V GPU card and the Italian MUR Departments of Excellence grant 2023-2027 “Quantum Frontiers”. MV and GMP acknowledge funding under project PNRR - Research infrastructures: Strengthening the Italian leadership in ELT and SKA (STILES). 
SL and GMP  acknowledge support by MUR under PRIN Project No. 2022FEXLYB, Quantum Reservoir Computing (QuReCo) and by the “National Centre for HPC, Big Data and Quantum Computing (HPC)” Project CN00000013 HyQELM – SPOKE 10.
We acknowledge the use of IBM Quantum Credits for this work. The views expressed are those of the authors and do not reflect the official policy or position of IBM or the IBM Quantum team.

\bibliography{bibliography}

\appendix
\section*{Appendix}

The work performed with IBM Fez would have not been possible without all the previous simulations performed with all the datasets shown in~\cref{fig:spectra}, which permitted to find the best setup to work with. In the course of~\cref{App:Training and Feature Test} we are going to review all the results obtained by the simulations from which we learned the minimal dimension of the training dataset and the number of principal components needed to solve the task successfully, while minimizing the quantum computational time needed to prove fault tolerance of the algorithm. On the other hand, in~\cref{App: Simulations Results} we show the result obtained by processing the 4 datasets shown in~\cref{fig:spectra} with the best setup possible deduced in~\cref{App:Training and Feature Test}.

\section{Training and Feature Tests}

Principal component analysis is an unsupervised learning method used to compress high-dimensional datasets. The compression of the input data set is a key feature to implement fault-tolerant quantum machine learning methods, but at the same time, it removes part of the information about the input. Therefore, it is important to identify a compromise between compression and minimal loss of information. 

\begin{figure}[h!]
    \centering
    \includegraphics[width=0.9\linewidth]{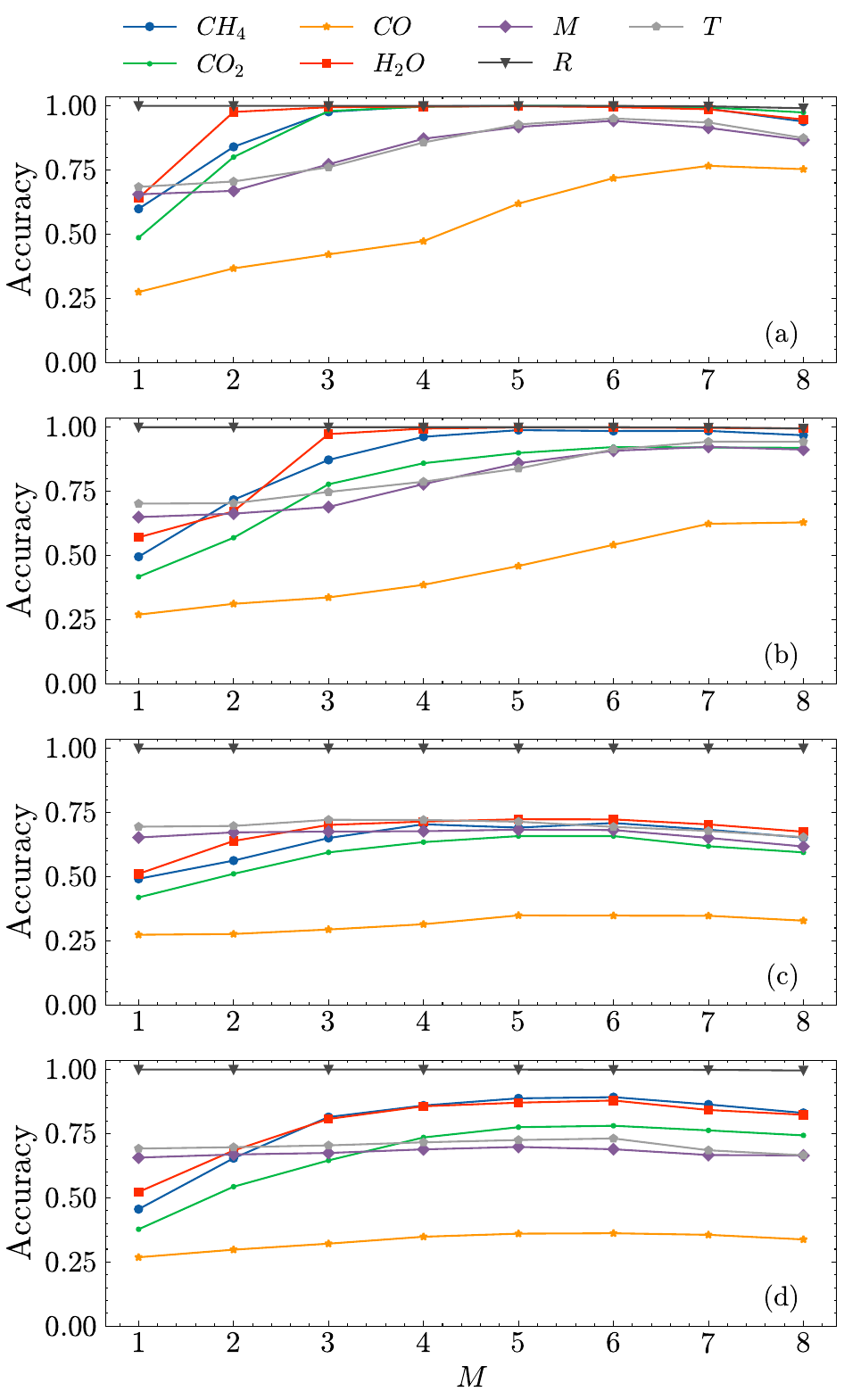}
    \caption{Accuracy of the algorithm varying the number of encoded principal components of the spectra. The simulations have been performed considering a dataset of $10^4$ spectra, with the $80\%$ used for training the remaining for testing. In the panel{\bf (a)} there are the performances using the {\bf TauREx} dataset, in the  {\bf (b)} there are those using the interpolated dataset in the spectral range of {\bf JWST}, in {\bf (c)} those with the dataset of {\bf NJWST} and in {\bf (d)} those with the dataset of {\bf FJWST}.}
    \label{fig:feature_test}
\end{figure}

In order to find this compromise, we trained the QELM by extracting a varying number of principal components from each patch of the spectra and plotted the accuracy of the corresponding QELM with fixed dimension of the training dataset. The {\bf NJWST} spectra were pre-processed at first as the {\bf TauREx} and {\bf JWST} dataset and secondly by applying a PCA filter to it. In particular, we computed the cumulative explained variance to get the number of principal components needed to de-noise the {\bf NJWST} dataset and reconstructed the filtered spectra ({\bf FJWST}~\cref{fig:spectra}) by retaining 10 components. The dimension of the training dataset was fixed to $D_{\text{train}} = 8000$ and we performed a test on $D_{\text{test}} = 2000$. The results of the feature test is visualized in ~\cref{fig:feature_test} where we observe that in the {\bf TauREx} and {\bf JWST} cases (\cref{fig:feature_test} (a) and (b)) the accuracy in the retrieval of all the atmospheric parameters eventually reach a plateau with $M {=} 7$ principal components; on the other hand, in the {\bf NJWST} case, the accuracy has an almost flat behavior for basically all of the parameters except for $CH_4, H_2O$ and $CO_2$ which grow slightly reaching a maximum for $M = 6$. The same maximum is achieved for $M = 6$ also in the {\bf FJWST} case where, contrary to the {\bf NJWST}, where the accuracy grows slightly. Furthermore, we can notice how the prediction on the radius is totally unaffected by the number of principal components, meaning that the information about it is not encoded in the principal components but in the global properties of the spectra. All the results are obtained at infinite statistics.

\begin{figure}[h!]
    \centering
    \includegraphics[width=0.9\linewidth]{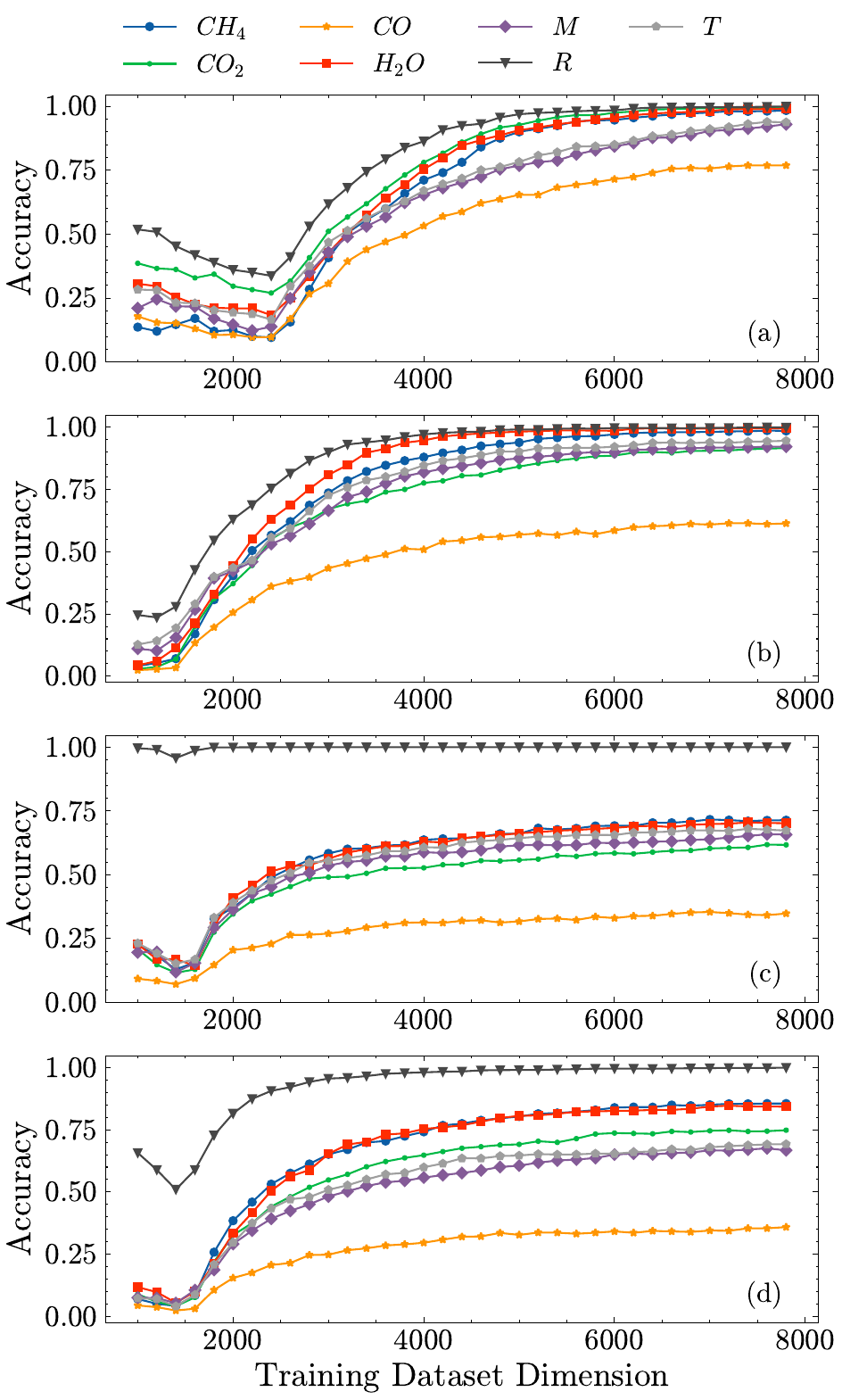}
    \caption{Accuracy of the algorithm varying the number of training states and considering a test of 2000 spectra. In the panel {\bf (a)} there are the performances using the {\bf TauREx} dataset, in {\bf (b)} there are those using the interpolated dataset in the spectral range of {\bf JWST}, in {\bf (c)} those with the dataset of {\bf NJWST} and in {\bf (d)} those with the dataset of {\bf FJWST}. The number of components used for each patch is $M {=} 7$ for the {\bf TauREx} and {\bf JWST} datasets while $M {=} 6$ for the {\bf NJWST} and {\bf FJWST} case.}
    \label{fig:training_test}
\end{figure}

The tests to find the minimal training dataset dimension were thus performed by considering $M {=} 7$ principal components for the {\bf TauREx} and {\bf JWST} datasets and $M {=} 6$ for the other two. The testing was again performed on $D_{test} {=} 2000$ spectra at infinite statistics. The results are visualized in~\cref{fig:training_test}, where we can observe that while with the {\bf TauREx} dataset the QELM needs about $D_{train} = 8000$ spectra to be fully trained, all the other datasets need only $D_{train} = 4000-5000$. This is due to the higher amount of patches and thus more information to be processed. This is translated also in a higher number of required reservoirs and therefore also outcomes, which result in a higher amount of parameters in the output layer.

\label{App:Training and Feature Test}
\section{Simulations Results}

The results obtained~\cref{App:Training and Feature Test} give us the opportunity to test the algorithm at its full potential at least on simulation.
\begin{figure*}
    \centering
    \includegraphics[width=\linewidth]{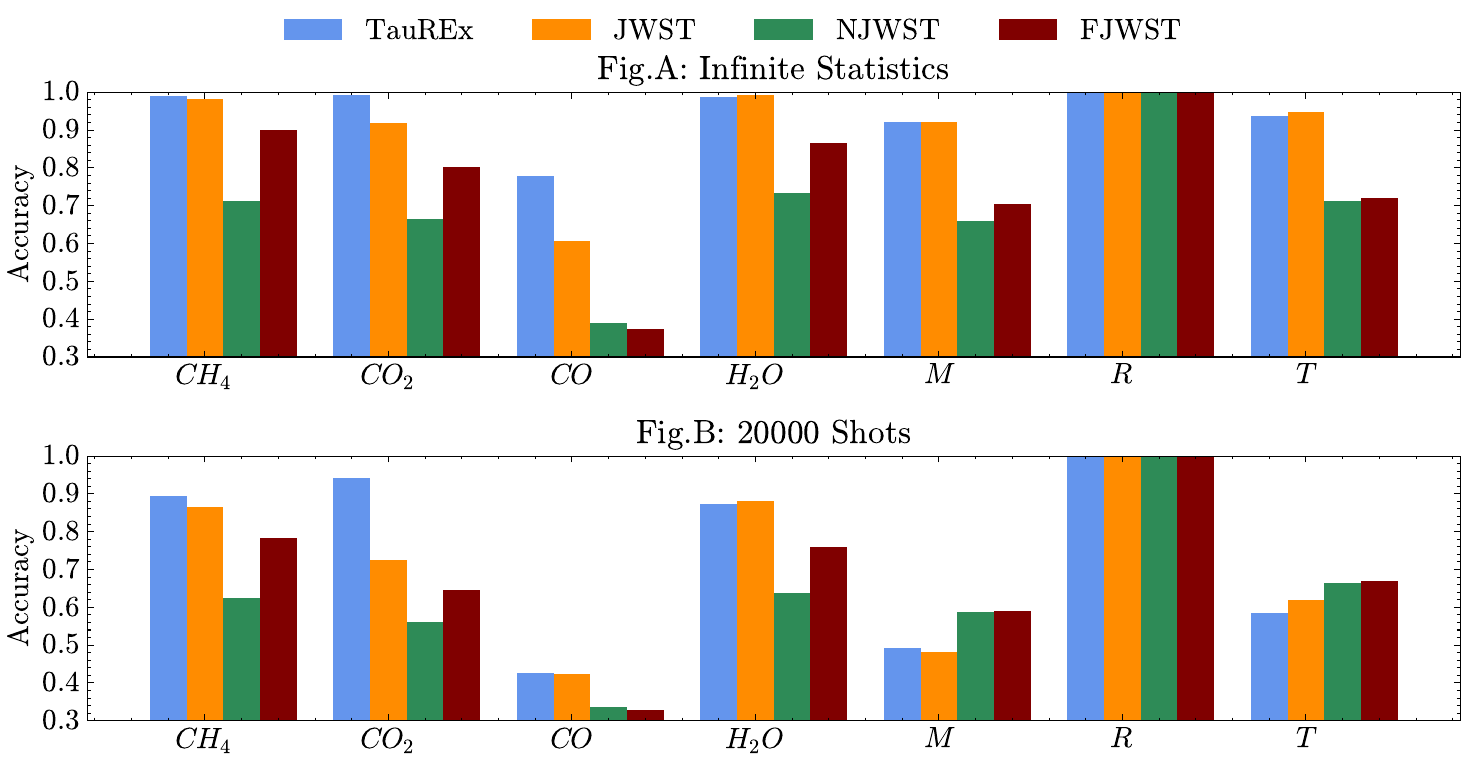}
    \caption{Accuracy of the QELM in retrieving each atmospheric parameter using the datasets shown in~\cref{fig:spectra}. In this context we used a dataset of $10000$ spectra split in $80-20\%$ respectively for training and testing and used $M = 7$ principal components for each patch for the {\bf TauREx} and {\bf JWST} datasets, while $M = 6$ for the other two. In {\bf Fig. A} are reported the results at infinite statistics, while in {\bf Fig. B} the results with a statistics of $20000$ shots.}
    \label{fig:Results}
\end{figure*}
\begin{figure*}
    \centering
    \includegraphics[width=\linewidth]{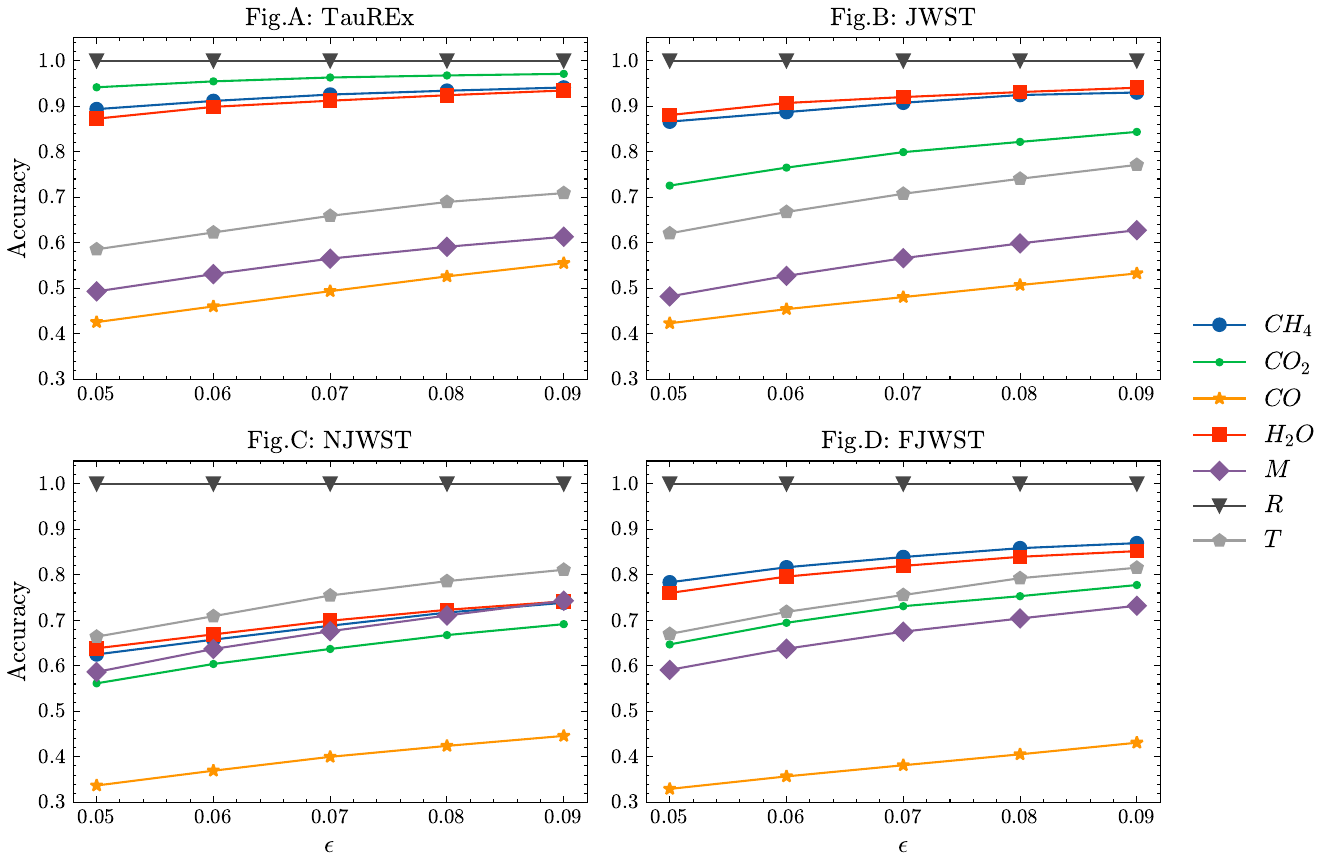}
    \caption{Accuracy of the QELM with 20000 shots varying the tolerance threshold indicating the successful rate defined in~\cref{eq:metric}.}
    \label{fig:tolerance}
\end{figure*}
In~\cref{fig:Results} are visualized the results obtained in the atmospheric retrieval considering the four types of dataset shown in~\cref{fig:spectra}. As can be seen from the histograms, in the {\bf TauREx} case we observe that basically all the parameters are retrievable at infinite statistics with an accuracy of $99-100\%$ except for the mass and temperature which still reach $93-94\%$ success rate. Furthermore, as already seen in~\cref{fig:hardware results}, also in this case $CO$ cannot be retrieved successfully neither at infinite statistics, meaning that its features are hidden by those of $CO_2$ which have similar spectral features. In the lower histogram of~\cref{fig:Results} we also observe that while we are able to retrieve $CH_4, CO_2, H_2O$ and $R$ with a success rate over the $87\%$, the same cannot be observed for mass and temperature which require a higher statistics to be estimated correctly.
In the {\bf JWST} case, the QELM shows a similar accuracy as in the {\bf TauREx} case, with just some slight worsening in the retrievability of $CO_2$ and $CO$, meaning that part of their spectral features were contained in the spectral range which was cut off.

Adding the shot noise to this dataset and processing it without a previously filtering the noise using the principal component analysis leads to a retrieval accuracy below the $60\%$ for all the parameters except for the radius which has robustness against noise in the retrieval since, as previously seen in~\cref{fig:feature_test}, its information is codified in the averages of the spectra and therefore it is not disturbed by the presence of shot noise. These performances can be enhanced by using a filtering procedure in the pre-processing. Indeed, without the filtering procedure, the the normalization process of each patch amplifies not only the spectral features but also the shot noise, invalidating the principal components collected before the encoding. As shown in~\cref{fig:Results}, the principal component filter improves the retrievals which have still a lower success rate than in the {\bf JWST} case, nonetheless perform better than the {\bf NJWST}, reaching a success rate of $65-78\%$ at finite statistics for $CH_4, CO_2$ and $H_2O$. Of course, all the results at finite statistics can be improved with a higher number of shots which should let us approach the infinite statistics results

In~\cref{fig:tolerance} we show how the accuracy on the retrieval depends on the tolerance as defined in~\cref{eq:metric}, in a finite statistics simulation with 20000 shots. From the slope of the functions and the absence of significant jumps, we can conclude that the tolerance chosen as threshold was not significant in the classification of the successful retrievals.

\begin{figure*}
    \centering
    \includegraphics[width = \linewidth]{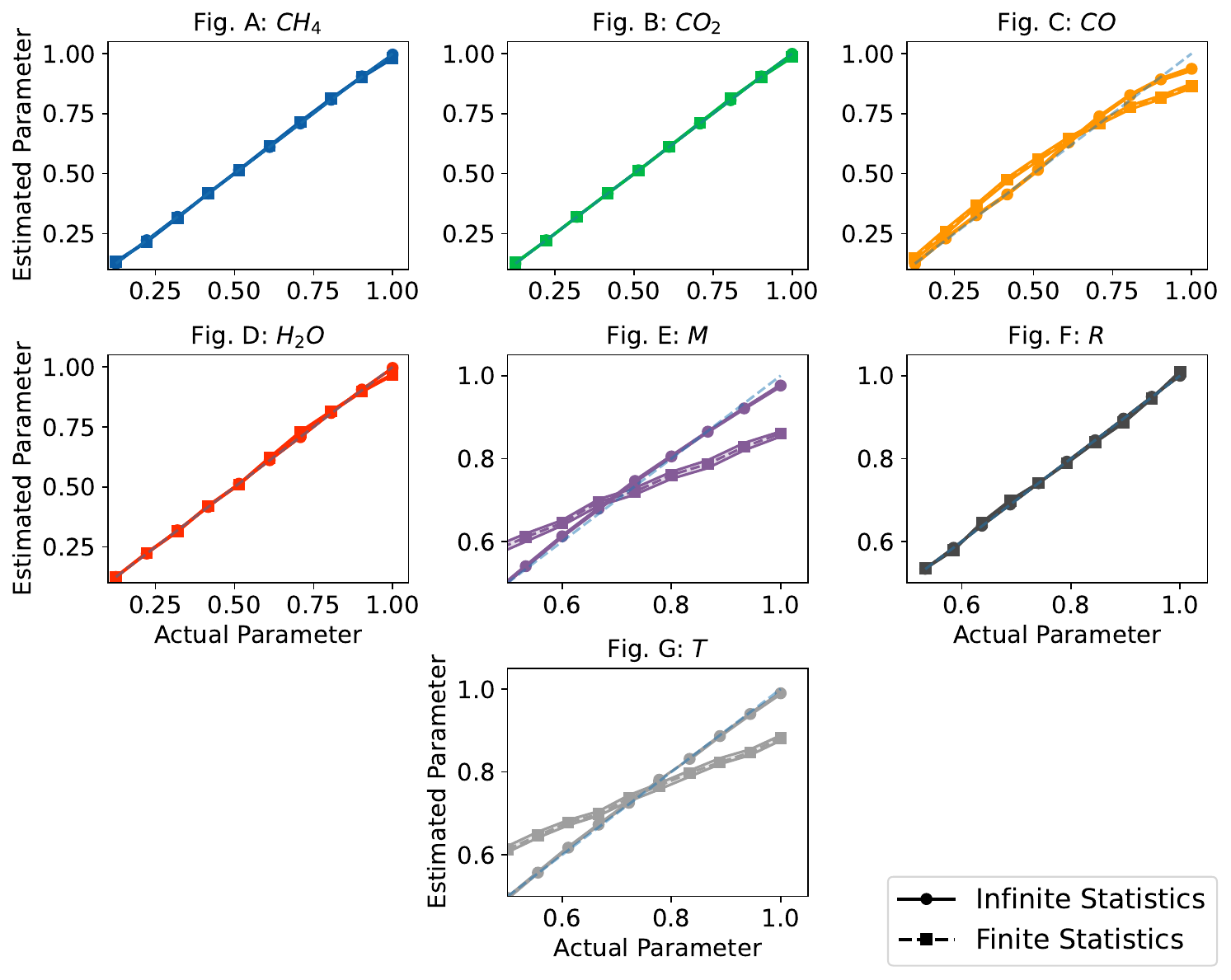}
    \caption{Comparison between the parameters as estimated with QELM and the real ones. The spectral data are generated using atmospheric parameters which have a discrete distribution in the intervals reported in~\cref{tab:parameters}. The estimation of each value has been obtained by making a bootstrap of the distributions obtained for each value. In the plot we show the lower and upper bounds of the confidence interval obtained with the bootstrap, with a confidence level of 95\%. The solid lines with marked with dots represent the estimation obtained with a infinite statistics simulation using the TauREx dataset with $M = 7$ and a dataset of $D = 10000$ spectra divided in $80-20\%$ among training and testing. The dashed lines marked with squares represent the processing of the same dataset made with a statistics of 20000 shots.}
    \label{fig:par_est_tau}
\end{figure*}

\begin{figure*}
    \centering
    \includegraphics[width = \linewidth]{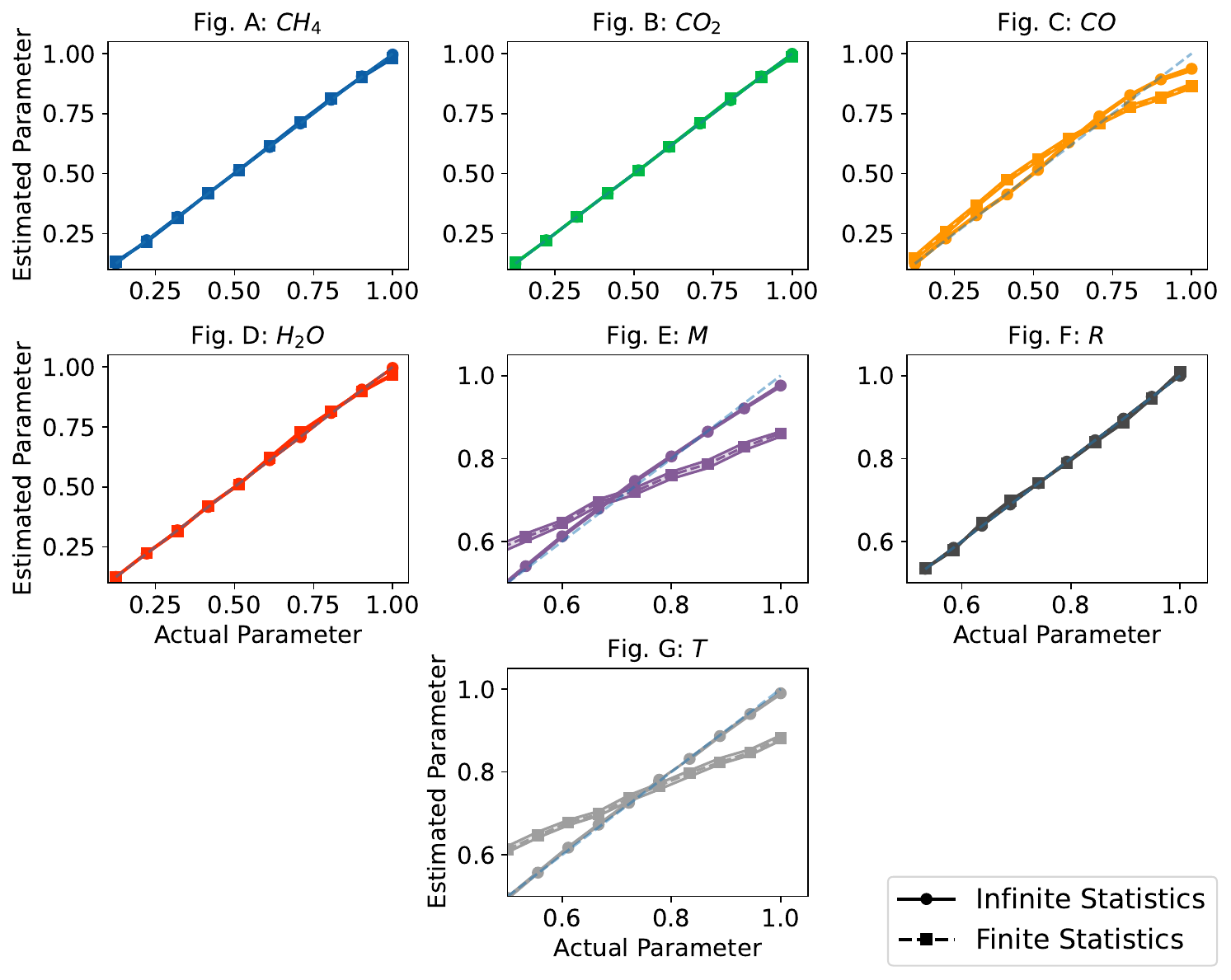}
    \caption{Comparison between the parameters as estimated with QELM and the real ones. The spectral data are generated using atmospheric parameters which have a discrete distribution in the intervals reported in~\cref{tab:parameters}. The estimation of each value has been obtained by making a bootstrap of the distributions obtained for each value. In the plot we show the lower and upper bounds of the confidence interval obtained with the bootstrap, with a confidence level of 95\%. The solid lines with marked with dots represent the estimation obtained with a infinite statistics simulation using the JWST dataset with $M = 7$ and a dataset of $D = 10000$ spectra divided in $80-20\%$ among training and testing. The dashed lines marked with squares represent the processing of the same dataset made with a statistics of 20000 shots.}
    \label{fig:par_est_jwst}

\end{figure*}

Finally in~\cref{fig:par_est_tau} and~\cref{fig:par_est_jwst} we observe the accuracy of the QELM in retrieving the atmospheric parameters depending on its value as previously done for the hardware case in~\cref{fig:parameter_estimation}. In particular, in~\cref{fig:par_est_tau} we see the results using the {\bf TauREx} dataset and in~\cref{fig:par_est_jwst} we observe those with the {\bf JWST} dataset. From the figures, it is possible to note how the precision in retrieving the highest values decreases, especially at finite statistics. Furthermore, we can note that, as in the hardware case (\cref{fig:parameter_estimation}), mass and temperature at finite statistics are very difficult to be retrieved, showing an higher demand of shots in order for the QELM to achieve better estimations.
\label{App: Simulations Results}

\end{document}